\documentclass[aps,prl,twocolumn,groupedaddress,epsf]{revtex4}

\usepackage{graphicx}
\usepackage{subfigure}

\usepackage{color}

\newcommand{\beq}{\begin{equation}}
\newcommand{\eeq}{\end{equation}}
\newcommand{\ov}{\overline}

\begin{document}

\title{Supersymmetric $U(1)_B \times U(1)_L$ model 
with \\ leptophilic and leptophobic cold dark matters }

\author{P. Ko}
\email[]{pko@kias.re.kr}
\affiliation{School of Physics, KIAS, Seoul 130-722, Korea}

\author{Y. Omura}
\email[]{omura@kias.re.kr}
\affiliation{School of Physics, KIAS, Seoul 130-722, Korea}

\date{\today}

\begin{abstract}
We consider a supersymmetric model with extra $U(1)_B \times U(1)_L$ 
gauge symmetry that are broken spontaneously.
Salient features of this model are that there are three different 
types of cold dark matter (CDM) candidates, and neutral scalar sector 
has a rich structure. Light CDM with  
$\sigma_{\rm SI} \sim 10^{-3\pm 1}$ pb can be easily accommodated 
by leptophobic dark matter ($\chi_B$) with correct relic density, if the $U(1)_B$ gauge 
boson mass is around $2 m_{\chi_B}$.
Also the PAMELA and Fermi/LAT data can be fit by leptophilic CDM with mass 
$\sim 1$ TeV. 
There could be interesting signatures of new fermions and 
new gauge bosons at the LHC. 
\end{abstract}

\maketitle
Proton decay is possible only if both baryon and lepton numbers
are broken. The current bound on proton life time indicates that 
both of these symmetries are extremely good ones. 
This fact might indicate that these symmetries are exact local symmetry, 
rather than global symmetries which are always expected to be broken 
to some extent.

Recently the gauged $U(1)_B \times U(1)_L$ model without 
supersymmetry has been constructed and phenomenological 
consequences of this model have been discussed \cite{wise}.
Anomaly can be cancelled by introducing new full family of 
the opposite (or the same) chiralities from the SM fermions
with the baryon number $-1 (1)$and lepton number $-3 (3)$,
respectively. Now, let us focus on the model with the opposite chirality.

Basic features of the model with a CDM candidate are additional
scalars, $S_L $ and $S_B$ that break the gauged $U(1)_L$ and 
$U(1)_B$ at some scale, and a new stable scalar $X$ which 
makes new heavy quark decay into light quark and $X$. 
The model also has interesting aspects for neutrino physics 
and leptogenesis \cite{wise}. 

The model considered in Ref.~\cite{wise}, being a nonsupersymemtric  
model with fundamental scalars, does not address the fine tuning 
problem of Higgs mass$^2$. 
If we consider the fine tuning problem seriously, we have to include 
another new physics at EW scale. 
Softly broken supersymmetry at EW scale is probably 
the simplest way to resolve this issue.
Furthermore, the lightest neutral particle can be stable in supersymmetric models with R-parity.  
The extension of  the gauged $U(1)_B \times U(1)_L$ model to SUSY can realize not only R-parity spontaneously,
but also additional CDM candidates. There is a possibility that we can provide several experimental results with a comprehensive explanation.

In this letter, we consider SUSY extension of $U(1)_B \times U(1)_L$ 
model with new full family fermions with opposite chirality, 
and study its phenomenological consequences.  (The model 
with the same chirality fermions is similar to the MSSM$_4$, and 
we do not pursue this possibility in this letter.) The matter content
of our model is similar to that of Ref.~\cite{wise}, but we add 
lepton number carrying superfield $X_L$ and $\overline{X_L}$,
which can make leptophilic CDM. 

Some of the salient features of our model are that both 
the dark matter and neutral Higgs  boson sector become 
very rich with multicomponent. There are three CDM candidates, leptophilic and leptophobic and the usual lightest 
neutralino. There are ten neutralinos and ten neutral scalars. 
After baryon and lepton numbers are broken spontaneously, 
there are mixing among new neutral fermions (scalars) and 
those in the minimal supersymmetric standard model (MSSM).  

In the SUSY $U(1)_B \times U(1)_L$ model, all gauge anomalies are 
cancelled by introducing one new family of quarks and leptons with
opposite chiralities from the SM fermions. (See Table.\ref{table1}.)
$S_B(\ov{S_B})$ and $S_L(\ov{S_L})$, whose charges are $n_B(-n_B)$ 
and $2(-2)$, are the fields which break $U(1)_B$ and $U(1)_L$.
In order to realize spontaneous $U(1)_B$ and $U(1)_L$ breaking, we set soft SUSY breaking terms
of $S_B(\ov{S_B})$ and $S_L(\ov{S_L})$, squared mass terms and B-terms, to ones in appropriate parameter regions like MSSM higgs without assuming
specific mediation mechanism.

$S_L$ and $\ov{S_L}$ couple with the right-handed neutrino, $N_i$, 
and generate heavy Majorana masses. 
$X_B$ and $\overline{X_B}$ are the fields which make extra quarks, 
such as $Q'$, decay. If we assume that superpotential is a polynomial, 
extra quarks cannot decay without $X_B$ and $\ov{X_B}$.
We forbid couplings between $S_B(\ov{S_B})$ and SM particles according 
to the charge assignment in Table \ref{table1}.   
Compared with non SUSY model of Ref.~\cite{wise}, we have additional
$X_L$ and $\overline{X_L}$ with $U(1)_L$ charges $\pm n_L$ which 
can be leptophilic CDM. 
In order to forbid $X_L$ coupling to leptonic bilinear, we impose 
$n_L \neq 0,~\pm 1,~\pm 2,~\pm 4$. The charge of  $S_B$, $n_B$, 
also satisfies $n_B \neq 0,~\pm 1/3,~\pm 2/3,~\pm 4/3$
to forbid direct couplings between $S_B(\ov{S_B})$ and $X_B(\ov{X_B})$ 
in renormalizable superpotential.   
For $n_L=\pm 1$ and $n_B=\pm 4/3$, operators like $S_L X_L^2$ and 
$S_B X_B^2$ are allowed and $X_L$ and $X_B$ could be stable, 
but we assume that $n_L \neq \pm 1$ and $n_B \neq\pm 4/3$  
for convenience.

Superpotential of the model is given by 
\begin{equation}
W = W_{\rm quark} + W_{\rm lepton} + W_{\rm bilinear}
\end{equation}
where the first two are Yukawa terms and the last one is the bilinear
term:
\begin{widetext}
\begin{eqnarray}
W_{\rm quark} & = & Y^u_{ij} Q^i U^j H_u + Y^d_{ij} Q^i D^j H_d 
+ Y^{'}_u Q^{'} U^{'} H_d + Y_d^{'} Q^{'} D^{'} H_u + 
\lambda_{Q i} X_B Q^{'} Q^i + \lambda_{U i} \overline{X_B} U^{'} U^i 
+ \lambda_{D i} \overline{X_B} D^{'} D^i 
\\
W_{\rm lepton} & = & Y^{l}_{ij} L^i E^j H_d + Y^{\nu}_{ij} L^i N^j H_u
+ Y_l^{'} L^{'} E^{'} H_u + Y_\nu^{'} L^{'} N^{'} H_d  \nonumber
\\
&&+ \lambda_i S_L L^{'} L_i + \lambda_{ij} N^i N^j S_L 
+ \lambda_{N i} N_i \overline{S_L} N^{'}+\lambda_{Ei} \ov{S_L} E^{'} E_i 
\\
W_{\rm bilinear} & = & \mu H_u H_d + \mu_{X_B} X_B \overline{X_B} 
+ \mu_B S_B \overline{S_B} + \mu_{X_L} X_L \overline{X_L} 
+ \mu_L S_L \overline{S_L}. 
\end{eqnarray}
\end{widetext}
In this model,  lepton and the baryon numbers are gauge charges, 
so that there are no usual $R-$parity violating couplings. 
Assuming that $n_B$ is $2k/3~(k \neq 0,~\pm\frac{1}{2},~\pm1,~\pm2)$, where $k$ is integer, 
$U(1)_B \times U(1)_L$ breaks to $Z_{2B} \times Z_{2L}$.
R-parity, $(-1)^{3B+L+2j}$, is automatically conserved 
after $U(1)_B \times U(1)_L$ symmetry breaking.

We assume that $X_L$ and $X_B$ do not get nonzero VEVs, 
in order to avoid large mixing between $Q'$ and $Q_i$ which 
would generate large FCNC.
In the lepton sector, the yukawa couplings, such as $ \lambda_i $, 
generate mass mixing because of nonzero $<S_L>$.
For example, $\mu \rightarrow e,~\gamma $ process gives 
strong constraints on the yukawa couplings,
\beq
|\lambda_{Ee} \lambda_{\mu } | \lesssim \frac{ m_{l'}^3}{\langle 
\ov{S_L} \rangle \langle S_L \rangle} \times \frac{10^{-13}}{GeV}. 
\eeq

\begin{table}
\caption{\label{table1}
$SU(3)_C \times SU(2)_L \times U(1)_Y \times 
U(1)_B \times U(1)_L$ charges of fields in this model. $i,j,k=1,2,3$ are 
the generation indexes for the SM fields, and the primed fields are newly 
added to cancel all the gauge anomalies. All fields are left chiral superfields.}
\begin{ruledtabular}
\begin{tabular}{|c|c||c|c|}
Field & Charges  & Fields & Charges
\\ \hline 
$Q^i$ &  $(3,2,1/6 ; 1/3,0) $ & 
$Q^{'}$ & $(\bar{3}, 2,-1/6 ; -1 , 0)$
\\
$U^i$ & $(\bar{3}, 1, -2/3 ; -1/3 , 0 )$ & 
$U^{'}$ & $(3,1,2/3 ; 1 , 0)$
\\
$D^i$ & $(\bar{3}, 1, +1/3 ; -1/3 , 0 )$ &
$D^{'}$ & $(3, 1, -1/3 ; 1,0)$
\\   \hline
$L^i$ & $(1, 2, -1/2 ; 0, 1)$ & 
$L^{'}$ & $(1,2,1/2 ; 0,-3)$ 
\\
$E^i$ & $(1, 1, 1 ; 0 , -1)$ &
$E^{'}$ & $(1,1,-1;0,3)$
\\
$N^i$ & $(1,1,0;0,-1)$ &
$N^{'}$ & $(1,1,0;0,3)$ 
\\   \hline
$H_u$ & $(1,2,1/2;0,0)$ & 
$H_d$ & $(1,2,-1/2;0,0)$
\\   \hline 
$X_B$ & $(1,1,0;2/3,0)$ & 
$\overline{X_B}$ & $(1,1,0; -2/3 , 0)$
\\   \hline
$S_L$ & $ (1,1,0;0,2)$ & 
$\overline{S_L}$ & $(1,1,0;0,-2)$
\\   \hline
$S_B$ & $(1,1,0; n_B , 0)$ &
$\overline{S_B}$ & $(1,1,0; - n_B , 0)$
\\   \hline 
$X_L$ & $(1,1,0; n_L ,0)$ & 
$\overline{X_L}$ & $(1,1,0; -n_L , 0)$
\\
\end{tabular}
\end{ruledtabular}
\end{table}

There are two charged scalars $H^\pm$ as in the MSSM. 
The neutral scalar bosons are much richer than the MSSM:
\begin{itemize}
\item CP-even : $(Re \left( S_B \right), Re \left( \ov{S_B} \right) ),
~(Re \left( S_L \right), Re \left( \ov{S_L} \right) )$
\item CP-odd :  $(Im \left( S_B \right), Im \left( \ov{S_B} \right) ),
~(Im \left( S_L \right), Im \left( \ov{S_L} \right) )$
\end{itemize}
After $U(1)_B \times U(1)_L$ is spontaneously broken, 
there are mixings between the MSSM neutral Higgs bosons 
and $L$ and $B$ charged neutral scalars ($S_B^0$, $S_L^0$
and their conjugates) at one-loop level.  
While MSSM Higgs bosons couple to the fermion masses, 
the $L$ or $B$ charged scalars couple to the baryon 
and lepton numbers of fermions. Therefore after making field redefinition
to the physical mass eigenstates, the neutral scalars can have decay 
patterns which are different from the MSSM Higgs bosons, especially 
for light fermions for which $m_f / v \sin\beta (\cos\beta) \sim g_B $ 
or  $g_L$. 
The mixing and upper bound for the lightest $U(1)_B$ higgs will be 
very small, like $O(10)$ GeV, in the parameter region we discuss below. 
We expect the lightest to evade constraints on colliders~\cite{Schael:2006cr}, 
but more detailed studies are in need~\cite{progress}.

The neutralino sector of our model is also enlarged due to the 
new gauge symmetries. Let us call $\lambda_B$ and $\lambda_L$ 
the gauginos of $U(1)_B$ and $U(1)_L$, respectively.
There are two more sets of neutral fermions, 
$(\lambda_B ,  \widetilde{S_B} , \widetilde{\overline{S_B}}) \equiv \Psi_B^{0T}$, 
and 
$(\lambda_L ,  \widetilde{S_L} , \widetilde{\overline{S_L}}) \equiv \Psi_L^{0T}$
in addtion to the usual 3 neutralinos
$( \widetilde{B} , \widetilde{W^3} , \widetilde{H_u} , \widetilde{H_d} ) 
\equiv \Psi_{GH}^{0T}$ (gauginos plus Higgsinos).
All in all there are 10 neutral Majorana fermions.
After the gauge symmetry of our model is spontaneously broken 
into $SU(3)_C \times U(1)_{\rm em}$, all these neutral fermions  
mix with each other either at one-loop or two loop level.  
The mass eigenstates $\chi_{n=1,2,..,10}^0$ of ten neutral fermions 
are linear combinations of $ \Psi_B^0$, $\Psi_L^0$ and $\Psi_{GH}^0$.

$\widetilde{X_L}(\widetilde{\ov{X_L}})$ and 
$\widetilde{X_B}(\widetilde{\ov{X_B}})$ are also neutral, 
but they do not have mixing with $\lambda_B$ according to 
the zero VEVs of the bosonic superpartners.
That is, they can be dark matters of Dirac fermion and 
respect accidental $U(1)$ symmetry. Let us denote the dirac 
fermions as just $\widetilde{X_B}$ and $\widetilde{X_L}$ 
in the following sections.
The bosonic sectors of $X_L$ and $X_B$ also have global $U(1)$, 
so that they cannot have mixing with the other neutral scalars, 
or the mixing could be very small.

Our model has multicomponent CDM's due to three conserved quantities:
the usual $R-$parity and two accidental global symmetries, $U(1)_{X_B}$ 
and $U(1)_{X_L}$:
\begin{widetext}
\begin{eqnarray}
U(1)_{X_B} : (X_B , \overline{X_B} , Q^{'} , U^{'} , D^{'} ) & \rightarrow & 
( e^{i \alpha} X_B , e^{- i \alpha} \overline{X_B} , e^{- i \alpha}Q^{'} ,
 e^{i \alpha} U^{'} ,e^{i \alpha} D^{'} )
\\
U(1)_{X_L} : ( X_L , \overline{X_L} )  & \rightarrow & 
( e^{i \beta} X_L , e^{- i \beta} \overline{X_L} )
\end{eqnarray}
\end{widetext}
The SM particles, extra quarks and leptons are $R-$parity even, 
whereas their superpartners and all gauginos are $R-$parity odd. 
While the extra neutral bosons, such as $S_B$, are R-parity even, 
their superpartner is odd.
R-parity realizes a stable particle in our model, so that 
$\chi^0_1$, $\widetilde{X_B}$, and $\widetilde{X_L}$ 
are good candidates for CDM.

Besides, both $X_B$ and $\widetilde{X_B}$ carry $U(1)_{X_B}$ 
charge $Q_{X_B} = +1$, and their conjugates carry $Q_{X_B} = -1$.  
For leptonic fields, $X_L , \widetilde{X_L}$ (and their conjugates) carry 
$Q_{X_L} = 1 (-1)$, whereas all the other particles carry no 
$U(1)_{X_L}$ charge.  There are 2 complex scalars charged under only 
$U(1)_{X_B}$, but they are not degenerate generally because of B-term, 
$B_X \ov{X_B}X_B$.
The heavier scalar field can decay to $2$ quarks and the lightest scalar. 
Unless the lightest is heavier than both $\widetilde{X_B}$ and $\chi_1^0$, 
it cannot decay because of the global symmetry.
The lightest $U(1)_{X_B}$ charged scalar particle, $X_{B1}$, 
can be leptophobic CDM. The $U(1)_{X_L}$ charged fields $X_{L1}$ (scalar) 
is also a good leptophilic CDM candidate.

Eventually, there are $1$ Majorana fermion, $2$ Dirac fermion, 
and $2$ complex scalar fields as good CDM candidates in this model.
Among them, $3$ particles can be stable corresponding to $3$ 
global symmetries, and the other $2$ fields decay to $2$ CDMs.
It depends on the mass spectrum which fields are stable or not, 
and each sets must allow extra quarks to decay.
The extra quark, such as $Q'$, is $R$-parity even and charged 
under $U(1)_{X_B}$, so that at least lighter $X_{B1}$ or $\widetilde{X_B}$ 
must be lighter than $Q'$.  If $X_{B1}$ is heavier, both of $\widetilde{X_B}$ 
and $\chi_1^0$ must be lighter than $Q'$ because of $R$-parity.

Generally speaking, $U(1)_{X_B}$ is anomalous and $U(1)_{X_B}$ breaking 
terms, such as $X_B S_B^{-2/3n_B} = X_B S_B^{-\frac{1}{k}} $ 
(with $|k| \neq 0, 1/2, 1, 2 $),  can be written down, including 
non-renormalizable terms.
However, such breaking terms allow mixing between $\widetilde{X_B}$ and 
neutralino, because $X_B$ gets nonzero VEV through tadpole terms like 
$X_B \langle S_B \rangle^{-\frac{1}{k}}$. 
If our superpotential is a polynomial in superfields, such dangerous 
operators would be forbidden  as far as $|k | > 1$ is satisfied, 
and the stability of $X_{B1}$ and dirac CDM, 
$\widetilde{X_B}$, could be guaranteed. 
This argument can be applied to $U(1)_{X_L}$, and $n_L$ cannot be 
even integer and $\pm1$ to respect $U(1)_{X_L}$.     

Since there are many new particles and parameters compared with the 
MSSM, it would be easy to imagine that there are rich phenomenology 
of this model both at colliders and in CDM.
In order to illustrate the richness of our model, we choose to fit both
(arguably) intriguing signature of light cold dark matter (lCDM) from direct detection 
experiments by CoGeNT~\cite{Aalseth:2010vx} and DAMA/LIBRA~\cite{Bernabei:2010mq}, and positron excess observed 
by PAMELA~\cite{Adriani:2008zr} and Fermi/LAT satellite~\cite{Abdo:2009zk}.  
We consider the scenario that there are one light leptophobic CDM, $\widetilde{X_B}$ or $X_{B1}$, lightest neutralino, and one heavy leptophilic CDM, $\widetilde{X_L}$ or $X_{L1}$.

CoGeNT and DAMA/LIBRA results are challenged by CDMS~
\cite{ Akerib} and 
XENON10~\cite{Angle}, so we need very careful analysis to realize them in our model~\cite{Schwetz:2010gv}. 
Even if they are excluded, 
there is a region close to the CoGeNT and DAMA signal region, where 
lCDM $\chi_B$ can have $\sigma_{\rm SI} \sim 10^{-3\pm 1}$ pb with an 
acceptable  thermal relic density, as long as $M_{Z_B}/g_B \sim 1$TeV and 
$M_{Z_B} \sim 2 m_{\chi_B}$  which is not excluded by collider data 
(see Fig.~1 and related discussions).

Keeping this remark in mind, let us focus on the leptophobic lCDM scenario 
around the CoGeNT and DAMA/LIBRA signal region.  
We assume that the lCDM claimed by CoGeNT and/or  DAMA/LIBRA
is a leptophobic dark matter, $\widetilde{X_{B}}~(X_{B1})$, in our model.

From the claimed cross section $\sigma_{SI} \sim 10^{-40}$ cm$^2$ and 
the lCDM mass $\sim$ 7 GeV~\cite{Hooper:2010uy}, 
we can obtain a constraint on a combination 
of $m_{Z_B}/g_B \sim 1$ TeV, 
if the contributions of superparticles are negligible because of their TeV-scale masses.  
Using this constraint, we can constrain further 
the mass of $m_{Z_B}$ by calculating $\Omega_{lCDM} h^2$ and assuming it is 
equal to or less than the WMAP measurement of $\Omega_{\rm CDM}$. 
We find that the relic density turns out too large for heavy $m_{Z_B}$.
We can use the $s-$channel annihilation on the $Z_B$ resonance in order to 
enhance the annihilation cross section, thereby getting the right amount of 
relic density for the cold dark matter.  Then there is more or less unique value
for $m_{Z_B} \sim 2 m_{lCDM} \sim 14$ GeV, with very small $U(1)_B$ gauge 
coupling $\alpha_B \sim 10^{-5}$. See Fig.~1 for detail. 
The $X_{B1}$ CDM does not have S-wave contributions, 
so that the allowed region is narrower.
Squark exchanging could also contribute to the relic density 
through the $t-$channel annihilation, but it would be difficult to enhance the annihilation cross section.

Such a light $U(1)_B$ gauge boson with a 
weak gauge coupling is definitely out of reach at the current searches from 
colliders and low energy processes \cite{murayama,murayama2}. 
It would be interesting and important to search for such light leptophobic gauge boson 
at the upcoming experiments.

\begin{figure}
\includegraphics[width=7cm]{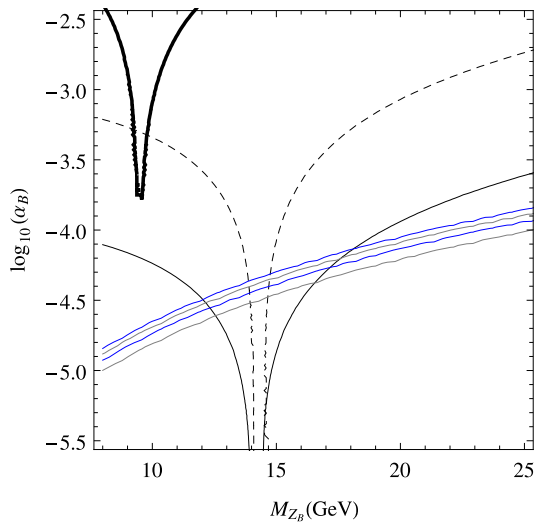}
\caption{\label{letterBL-fig0} Allowed region in the 
$(\log \alpha_B , M_{Z_B} )$ plane for $m_{lCDM} = 7$ GeV, 
that is consistent with WMAP data, is the inside of each 2 line, 
where the leptophobic lCDM is  bosonic $X_{B1}$ (dashed lines) 
or fermionic $\widetilde{X_B}$ (solid lines).  
Inside of thick black line is excluded by hadronic decay of $\Upsilon (1S)$. 
The bands within two blue lines and two gray lines are DAMA/LIBRA and 
CoGeNT regions \cite{Hooper:2010uy}.  }
\end{figure}

\begin{figure}
\includegraphics[width=7.5cm]{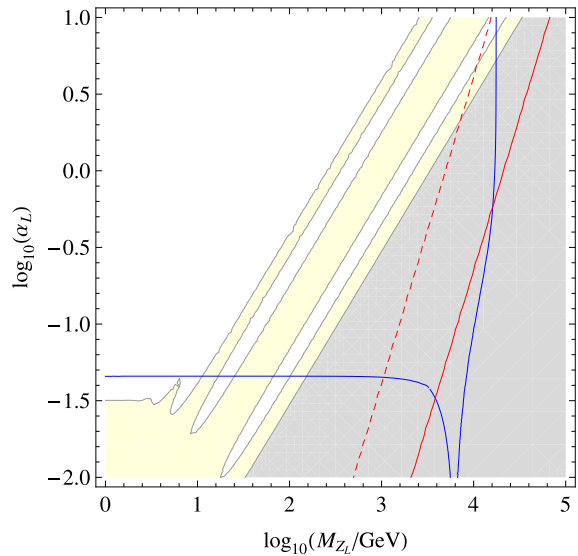}
\caption{\label{plotZL-3TeV} This figure shows the relic density and
Sommerfelt enhancement of $\widetilde{X_L}$ with $m_{CDM}=3$ TeV 
and $Q_{X_L}=1$. The above of the blue line corresponds to $(\Omega h^2 )_{X_L} \le 0.11$. In the white,
yellow and gray regions, the enhancements are $S \ge100$, $100 \ge S \ge 10$, and $10 \ge S$. 
The left sides of the dashed red and the red lines are exculded by $e,~\nu_{\mu}$ scattering and LEP-I\hspace{-.1em}I bound.
 }
\end{figure}

Positron excess observed by PAMELA collaboration~\cite{Adriani:2008zr} 
might be due to dark matter pair annihilation into charged lepton pairs.  
There are explicit models based on leptophilic $U(1)$ symmetries 
\cite{fox, baek}. 
Since our model has lepton charged carrying neutral scalar ($X_{L1}$) or 
fermion ($\widetilde{X_L}$) that couples only to leptons due to the gauged 
$U(1)_L$ and supersymmetry, we could try to fit PAMELA data using 
this leptophilic dark matter. 

In order to realize PAMELA, we have to consider large annihilation cross section,
compared with $\langle \sigma v \rangle \sim 1$ pb which is favored by WMAP data, $\Omega h^2 ~ \sim 0.11$.
In our model, its pair annhilation channels will be 
\[
\tilde{X_L} \tilde{X_L^\dagger} \rightarrow Z_L^* \rightarrow l l , 
{\rm or} ~ Z_L Z_L
\rightarrow 4 l's.  
\]
The latter is dominant if $Z_L$ is lighter than $\widetilde{X_L}$. 
In that case, Sommerfeld enhancement could work at low energy like
$\langle \sigma v (T_0) \rangle \sim S \langle \sigma v  (T_0) \rangle_0$.
In the Fig.~\ref{plotZL-3TeV}, we can see the enhancement and the relic density with $m_{CDM}=3$TeV.
However, $g_L$ and $M_{Z_L}$ are strongly constrained by the experiments: $e,~\nu_{\mu}$ scattering constrains 
$M_{Z_L}/g_L$ as $M_{Z_L}/g_L  \gtrsim 1.4$ TeV~\cite{Hagiwara:1994pw},
and the LEP-I\hspace{-.1em}I bound, $M_{Z_L}/g_L  \gtrsim 6$ TeV, which is discussed in $U(1)_{B-L}$, could be applied to our model \cite{Carena:2004xs}.

We notice that it is difficult to find the favored region with large $S$ according to Fig. \ref{plotZL-3TeV}.
The heavier $\widetilde{X_L}$ with large $S$ could avoid the constraints, but such heavy $\widetilde{X_L}$ pair annihilations may produce high energy neutrinos (upto the mass of parent $\widetilde{X_L}$) of 3 flavors at the same rates, which may cause conflict with the cosmic observation~\cite{baek}.

Another CDM also contributes to the relic density, but if it has a heavy mass, 
like $100$GeV,  the contribution could be enough small, as several works 
have discussed so far.
For example, it has been discussed that yukawa coupling, $\lambda_b X_{B1}b'b$, 
can realize not only small relic density of $X_{B1}$ but also the favored direct 
scattering cross section in Ref.~\cite{Feng:2008dz}.
If $\chi_1^0$ is MSSM-like CDM, $Z$ boson and Higgs exchanging are 
helpful \cite{Jungman:1995df}.

Kinematic mixing terms among $U(1)_B$, $U(1)_L$ and $U(1)_Y$ are allowed 
by those gauge symmetries, and generated radiatively.  
Their upper bounds are very tight~\cite{murayama2}, but the radiative 
correction is very tiny if gauge couplings are small enough as in our scenario. 
In any case, the initial condition must be controlled to suppress the kinematic 
mixing, as discussed in Ref.~\cite{murayama2}.

The conserved quantities are $U(1)_{X_B}$, $U(1)_{X_L}$ charges and 
$R-$parity. Note that $X_B$ and $X_L$ are $R-$parity even, whereas 
their fermionic superpartners are $R-$parity odd. 
Let us discuss some decays of new particles as examples ($\chi_1^0$ is the 
lightest $R-$):
\begin{eqnarray}
\lambda_B & \rightarrow & 
q \widetilde{q^*} \rightarrow q ( q \chi_1^0 ) 
\\
X_B & \rightarrow & q \bar{q} \chi_1^0 \widetilde{X_B}
\end{eqnarray}
$\widetilde{S_B}$ and its conjugate mix with $\lambda_B$ and 
decay into $q \widetilde{q^*} \rightarrow q ( q \chi_1^0 )$. 
Similarly, $\lambda_L$, $\widetilde{S_L}$ and $X_L$ 
decay into $l \bar{l} \chi_1^0 \widetilde{X_L}$.
Unlike the MSSM, the $R-$parity even scalar $X_B$ can decay into 
$q \bar{q} \chi_1^0 \widetilde{X_B}$ through two processes:  
\begin{eqnarray*}
 X_B & \rightarrow & \psi_{X_B} \lambda_B \rightarrow 
 \widetilde{X_B} ( q\bar{q} \chi_1^0 )
\\ 
& \rightarrow & \widetilde{Q^{'}} \widetilde{q_i^*} 
\rightarrow ( q \widetilde{X_B} )  
(q_j \chi_1^0 ) 
\end{eqnarray*} 
if kinematically allowed.  However $X_B$ can be stable if this decay is 
kinematically forbidden.  This is guaranteed by conservations of 
$U(1)_{X_B}$, $U(1)_{X_L}$ charges and $R-$parity.
If $\widetilde{X_B}$ is heavier than $X_B$, the reverse process will occur 
if kinematically allowed.  The new heavy (s)quarks can be produced at
the LHC: $pp \rightarrow t^{'} \overline{t^{'}} , \widetilde{t^{'}} 
\widetilde{\overline{t^{'}}}.$
If $X_B$ is lighter than $\widetilde{X_B}$,  the main decay modes of 
$t^{'}$ and $\widetilde{t^{'}}$ are $t^{'} \rightarrow q X$ and 
$\widetilde{t^{'}} \rightarrow X_B \widetilde{q} \rightarrow 
X_B q \chi_1^0$, respectively.
If $\widetilde{X_B}$ is lighter than $X_B$, then the main decay modes of
$t^{'}$ and $\widetilde{t^{'}}$ are $t^{'} \rightarrow \widetilde{X_B} 
\widetilde{q} \rightarrow \widetilde{X_B}  q \chi_1^0$, and  
$\widetilde{t^{'}} \rightarrow \widetilde{X_B} q$. 
If some of the decays are kinematically forbidden, 
$t^{'}$ or $\widetilde{t^{'}}$ will be stable and hadronized, 
and thus strongly constrained by heavy charged particle search.
Finally, new scalars $S_B$, $S_L$ and their conjugates will mix 
with neutral Higgs and will decay into lighter fermions or weak 
gauge bosons or $Z_B$'s. 

There are strong constraints on $t^{'}$ and $b^{'}$ masses. 
They can decay to quarks and CDM,  but such exotic quarks search 
could be replaced by SUSY particle ones, such as squark decay process, 
$\widetilde{t} \rightarrow \chi_1^0,~t$.
In Ref.~\cite{Alwall:2010jc}, the bounds for $t^{'}$ and $b^{'}$ masses, 
which  correspond to the case with scalar $X_B$ CDM, are dicussed 
in that way: the lower masses are more than $300$GeV.
If this argument is applied to our model, Landau poles of Yukawa 
couplings appear at TeV scale. 

We must also care about EWPT bound for higgs mass, extra quark masses, 
and all sparparticle masses.
The discussions in Ref.~\cite{Kribs:2007nz} can be applied if superparticle 
and extra higgs contributions are small.

In this letter, we constructed supersymmetric extension of $U(1)_B \times U(1)_L$
model, extending the works by Wise et al. \cite{wise}. The model is anomaly free, 
with very rich phenomenology in terms of colliders, Higgs bosons, flavor physics 
and CP violation, and most notably CDM.  The most interesting aspect of this 
model is the multi component CDM's. The model houses both leptophilic and 
leptophobic CDM's in addition to the usual lightest neutralino, 
and the stability of these CDM's are consequence of gauge symmetry and particle 
contents, not of {\it ad hoc} $Z_2$ symmetry. 
Furthermore, the leptophobic CDM can be light enough to accommodate 
the CoGeNT/DAMA signal region, with correct amount of relic density.  
The leptophilic CDM can be heavy enough to explain the positron excess 
observed by PAMELA and Fermi/LAT. 
However, it seems to be difficult to explain PAMELA without assumption of large boost factor. 

We could describe the simpler scenario that the lightest neutralino approximately becomes leptophilic because of small mixing, and realizes PAMELA data. Furthermore, our model has an interesting signature that the missing $E_T$ 
signature could be due to two different DM (with different masses and spins).
We touched relevant features of our model with a wide brush, relegating the 
details of phenomenology to the future works \cite{progress}.

\begin{acknowledgments}
We thank to P. Gondolo and M. Drees for useful discussions.
This work is supported in part by SRC program of National Research 
Foundation (NRF) through Korean Neutrino Reactor Center (KNRC), 
Seoul National University. 
\end{acknowledgments}

\end{document}